# Unified Growth Theory Contradicted by the Economic Growth in the Former USSR

Ron W Nielsen[1]

Environmental Futures Research Institute, Gold Coast Campus, Griffith University, Qld, 4222, Australia

December, 2015

Historical economic growth in the countries of the former USSR is analysed. It is shown that Unified Growth Theory is contradicted by the data, which were used, but not analysed, during the formulation of this theory. Unified Growth Theory does not explain the mechanism of economic growth. It explains the mechanism of Malthusian stagnation, which did not exist and it explains the mechanism of the transition from stagnation to growth that did not happen. Unified Growth Theory is full of stories but it is hard to decide which of them are reliable because they are based on unprofessional examination of data. The data show that the economic growth in the former USSR was never stagnant but hyperbolic. Industrial Revolution did not boost the economic growth in the former USSR. Unified Growth Theory needs to be revised or replaced by a reliable theory to reconcile it with data and to avoid creating the unwarranted sense of security about the current economic growth.

**Introduction**

We have already demonstrated that the Unified Growth Theory (Galor, 2005a, 2011) is repeatedly contradicted by data (Nielsen, 2014, 2015a, 2015b, 2015c, 2015d). It is contradicted by the world economic growth and by the economic growth in Western Europe (Nielsen, 2014). It is contradicted by the GDP/cap data (Nielsen, 2015a), by the economic growth in Africa (Nielsen, 2015b), by the economic growth in Asia (Nielsen, 2015d) and implicitly by the mathematical analysis of the historical economic growth showing repeatedly that global, regional and national economic growth was hyperbolic (Nielsen, 2015c).

We have demonstrated that Unified Growth Theory is contradicted by the same data (Maddison, 2001) that were used during its development. Unfortunately, this excellent data, published by the well-known economist were never analysed by Galor. Conclusions about the mechanism of economic growth were based on the generally-accepted doctrines and on presenting Maddison's data in a grossly simplified and misleading manner (Ashraf, 2009; Galor, 2005a, 2005b, 2007, 2008a, 2008b, 2008c, 2010, 2011, 2012a, 2012b, 2012c; Galor and Moav, 2002; Snowdon & Galor, 2008). Such a way of using data is bound to lead to incorrect conclusions.

---

[1]AKA Jan Nurzynski, r.nielsen@griffith.edu.au; ronwnielsen@gmail.com;
http://home.iprimus.com.au/nielsens/ronnielsen.html

In our discussion we shall use the latest data describing economic growth (Maddison, 2010). In this publication, Maddison extended the data to 2008, but any of his publications can be used to demonstrate that the Unified Growth Theory is contradicted by data.

We shall focus our attention on the economic growth in the countries of the former USSR and we shall test Galor's fundamental postulate of the existence of the three, distinctly different, regimes of growth: the Malthusian regime of stagnation, the post-Malthusian regime and the sustained-growth regime. The concept of these three regimes of growth was explained by Galor (2005a, 2008, 2011, 2012a). Briefly, Malthusian regime of stagnation was supposed to have commenced in 100,000 BC and lasted until 1750 for developed countries. Countries of the former USSR are in this group (BBC, 2014; Pereira, 2011). The post-Malthusian regime was allegedly between 1750 and 1870, and the sustained-growth regime from 1870. The post-Malthusian regime overlaps the Industrial Revolution, 1760-1840 (Floud & McCloskey, 1994). For developing countries, the Malthusian regime was supposed to have ended in 1900.

The timing of the three regimes of growth is also based on the customary crude examination of data (see Galor, 2005a, p. 187). One incorrect step leads to another and soon the whole array of "Mysteries of the growth process" (Galor, 2005a, p. 220) are created, which need to be explained, including the mystery of the differential takeoffs, the mystery of the stunning escape from the Malthusian trap and the "mind-boggling phenomenon of the Great Divergence" (Galor, 2005, p. 220), all such features contradicted by the scientific analysis of the same data.

Galor is puzzled by the strange behaviour of the GDP/cap distributions. These distributions are hard to understand but their analysis can be made simple and their strange features easily explained (Nielsen, 2015a). They are not made of two or three different components governed by different mechanisms of growth, and the apparent take-off is just an illusion. The economic growth *was* slow over a long time and fast over a short time but is *impossible* to determine the transition from the slow to the fast growth because these distributions increase *monotonically*. The GDP/cap distributions have to be explained as a whole and the same mechanism has to be applied to the slow and the fast growth.

Galor imagines a spectacular transition from stagnation to growth. He describes it as a "remarkable escape from the Malthusian epoch" (Galor, 2005a, p. 177) or as "the stunning recent escape from the Malthusian trap" (Galor, 2005a, p. 220). We have already demonstrated (Nielsen, 2014, 2015a, 2015b, 2015c, 2015d) that there was no escape because there was no trap. The economic growth was unconstrained during the alleged but non-existent epoch of stagnation. Descriptions of stagnation and escapes from stagnation to growth are stories based on the incorrect interpretations of data, stories contradicted repeatedly by Maddison's data (Maddison, 2001, 2010).

We shall now demonstrate that the same conclusions apply also to the economic growth in countries of the former USSR. We shall show that at the time of the alleged remarkable or stunning escape from Malthusian trap, economic growth in these countries was increasing along the totally undisturbed hyperbolic trajectory, remarkably contradicting the concepts of stagnation, Malthusian trap and the escape from this trap.

As before, we shall use two ways of displaying data: (1) semilogarithmic display and (2) the display of the reciprocal values.

Hyperbolic growth is described by the following simple mathematical formula:

$$S(t) = (a - kt)^{-1} \qquad (1)$$

where, in our case, $S(t)$ is the GDP while $a$ and $k$ are positive constants.

The reciprocal of the hyperbolic distribution is a straight line:

$$\frac{1}{S(t)} = a - kt \qquad (2)$$

The advantage of using semilogarithmic displays is that they allow for an easy examination of data varying over a large range of values. The added advantage of using the reciprocal values of data is that they can help in identifying even small deviations from hyperbolic distributions. Galor's "remarkable" or "stunning" escape from Malthusian trap would be readily identified by a clear change in the trajectory fitting the reciprocal values of data.

If the reciprocal values of data follow a decreasing straight line, the growth is not stagnant but hyperbolic. If the growth is boosted, the reciprocal values are diverted to a steeper trajectory. If the growth is slowed down, the reciprocal values of data are diverted to a less steep trajectory. If the straight line remains unchanged, then obviously there is no change in the mechanism of growth. It makes no sense to divide a straight line into two or three arbitrarily selected sections and claim different regimes of growth controlled by different mechanisms.

**Analysis of data for the former USSR**

Economic growth in the countries of the former USSR between AD 1 and 2008 is presented in Figure 1. Reciprocal values of the GDP data, 1/GDP, are shown in Figures 2 and 3. The growth was hyperbolic between AD 1 and around 1870. Parameters describing hyperbolic fit to the data are $a = 6.547 \times 10^{-1}$ and $k = 3.452 \times 10^{-4}$.

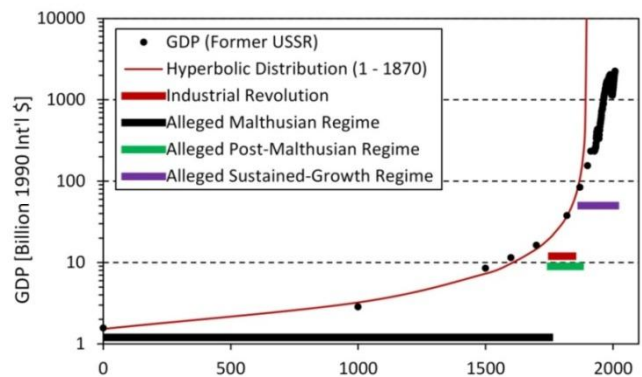

**Figure 1**. Economic growth in the countries of the former USSR between AD 1 and 2008, as represented by Maddison's data (Maddison, 2010), is compared with the hyperbolic distribution and with the unsubstantiated interpretations of the mechanism of growth proposed by Galor (Galor, 2005a, 2008a, 2011, 2012a). The alleged Malthusian regime of stagnation did not exist and neither did the alleged post-Malthusian and sustained-growth regimes. The Industrial Revolution had absolutely no impact on changing the economic growth trajectory. There was also no dramatic transition to a new and faster economic growth after the alleged epoch of stagnation, no transition from stagnation to growth at any time because there was no stagnation and no escape from the Malthusian trap because there was no trap. In place of these imaginary features there was the undisturbed and well-sustained hyperbolic growth. During the alleged sustained-growth regime, which was supposed to be characterised by the fast-increasing growth after the epoch of stagnation, the data show a diversion to a *slower* trajectory after the fast-increasing hyperbolic growth. The GDP is in billions of 1990 International Geary-Khamis dollars.

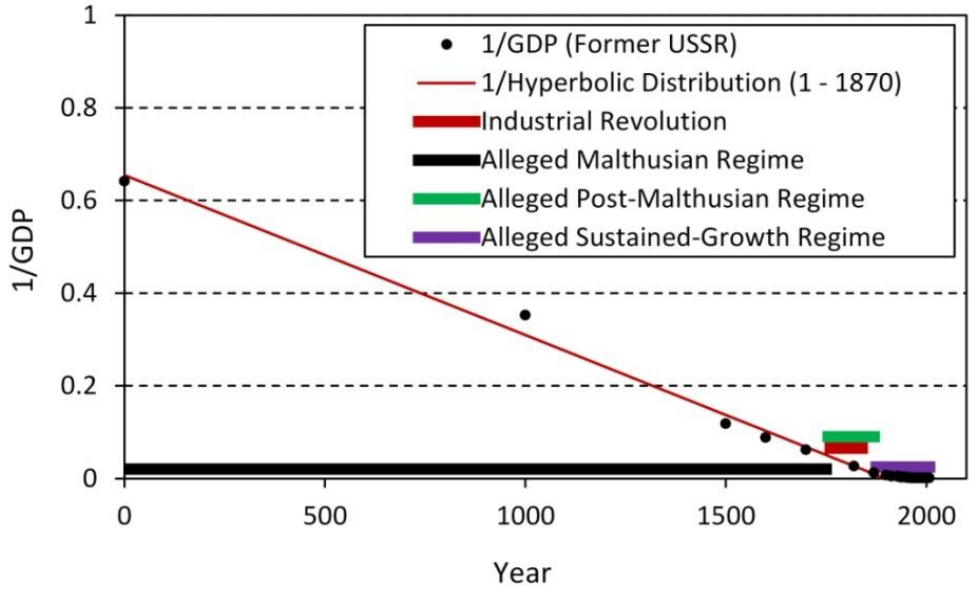

**Figure 2.** Reciprocal values of the GDP data, 1/GDP, for the former USSR are compared with the hyperbolic distribution represented by the decreasing straight line. Industrial Revolution did not boost the economic growth. The alleged Malthusian regime of stagnation did not exist and there was no transition from stagnation to growth at any time.

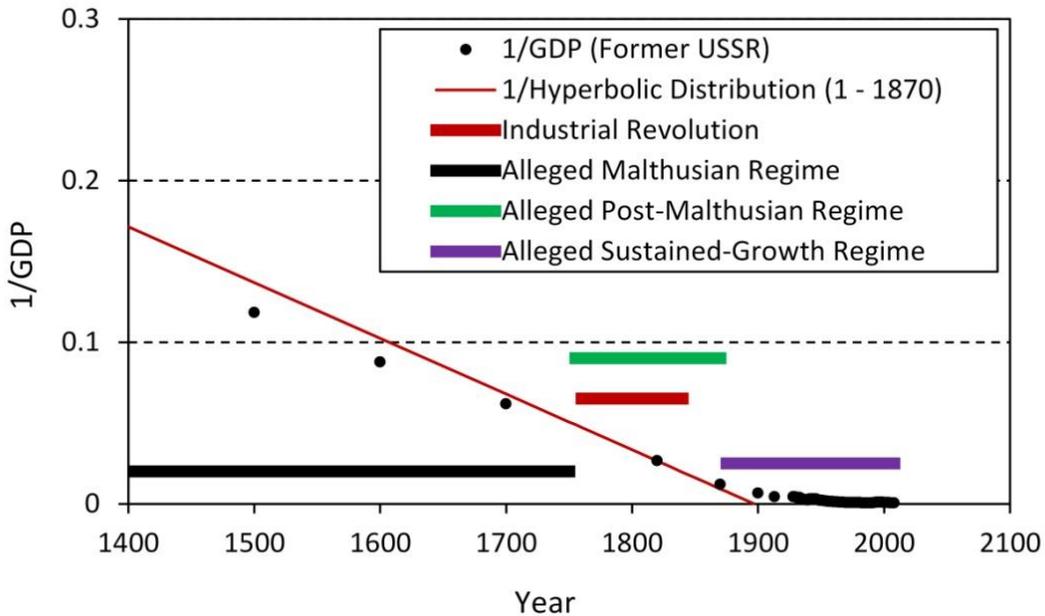

**Figure 3.** The end part of the plot of the reciprocal values of the GDP data, 1/GDP, for the former USSR. Economic growth was hyperbolic until around 1870 when it started to be diverted to a slower trajectory indicated by an upward bending of the reciprocal values trajectory. Industrial Revolution did not boost the economic growth. The alleged Malthusian regime of stagnation did not exist and there was no transition from stagnation to growth at any time because there was no stagnation. The "stunning" or "remarkable" escape from Malthusian trap (Galor, 2005a, pp. 177, 220) did not happen because there was no trap.

The epoch of Malthusian stagnation did not exist. Galor's regimes of growth are hanging there without having any connection with data. The "remarkable" or "stunning" escape from Malthusian trap did not happen because there was no trap. Galor's Malthusian regime ends in the middle of nowhere. Absolutely nothing (remarkable or less-remarkable, stunning or less stunning) happened on the border between the alleged Malthusian regime and the post-Malthusian regime. There was also no stunning or remarkable escape at the onset of the alleged sustained-growth regime. There was no dramatic increase in the economic growth. On the contrary, economic growth was diverted to a slower trajectory.

What is remarkable about the confrontation of Galor's theory with the empirical evidence is that there is such a consistently repeated and stunning disagreement between his theory and the data. The data also demonstrate that the Industrial Revolution had absolutely no impact on changing the economic growth trajectory in the countries of the former USSR.

According to Galor, Industrial Revolution was "the prime engine of economic growth" (Galor, 2005a, p. 212). It certainly was not in the countries of the former USSR and it certainly was not in Western Europe (Nielsen, 2014). This engine did not propel the economic growth along a new and faster trajectory. Whatever the engine was doing, it certainly did not boost the economic growth. These and other examples we have already investigated (Nielsen, 2014, 2015a, 2015b, 2015c, 2015d) show that exciting stories can be created and published even in academic journals and books but such stories have to be confronted by data because there is no room for fiction in scientific research.

The data and their analysis give no support for the concept of Malthusian stagnation and for the assumption of the existence of the steady-state Malthusian equilibrium. Economic growth was increasing along a remarkably-stable hyperbolic trajectory. There was no escape from Malthusian trap, let alone a "remarkable" or "stunning" escape as claimed by Galor (2005a, pp. 177, 220), because there was no trap. The growth was always unconstrained because the hyperbolic trajectory remained unimpeded.

The concept of stagnation is dramatically contradicted by data and so is the alleged transition from stagnation to growth. Such a transition never happened. On the contrary, from around 1870, economic growth in the countries of former USSR started to be diverted to a slower trajectory, away from its faster, historical hyperbolic trajectory.

Industrial Revolution did not boost economic growth. There is not even a slightest indication in the data that the Industrial Revolution had any effect on the economic growth trajectory. It is as if this event had never happened. Industrial Revolution might have had other effects on the lives of people but it did not boost the economic growth.

If not for Maddison and his data, the established knowledge in the economic research would have remained established, but now it has to be revaluated and changed. However, new insights should be welcome, particularly if they suggest a simpler explanation of the historical economic growth.

New insights are not only welcome in science but they are even searched for, because they contribute to a better understanding of studied phenomena and they also open new lines of investigations. There is no reason why new insights should not be welcome in the economic research unless the established doctrines are accepted by faith. In such a case, they have to be emotionally guarded and defended.

**Summary and conclusions**

Unified Growth Theory contains many stories and explanations but it is hard to decide which of them are relevant or even which of them have any scientific merit; probably not many because they are based firmly on the unprofessional examination of data. They seem to have little or no relevance to explaining the mechanism of the economic growth because there is no convincing connection between these stories and the data (Maddison, 2001, 2010).

Constructing numerous complicated arithmetical formulae does not make a theory reliable and acceptable. Incorrect ideas remain incorrect even if translated into a mathematical language. Concepts have to be tested by data and the Unified Growth Theory repeatedly fails the test (Nielsen, 2014, 2015a, 2015b, 2015c, 2015d).

Unified Growth Theory is based on the unsubstantiated postulate of the existence of the distinctly-different regimes of growth. It is also based on two other unsubstantiated postulates: the postulate of the differential takeoffs and the postulate of the great divergence. In due time we shall examine both of them and we shall demonstrate that they are also contradicted by data. Under these conditions it is hard to know how much can be rescued from this theory. A new approach to the explanation of the mechanism of the historical economic growth is needed.

Unified Growth Theory does not explain the mechanism of economic growth. It explains features that do not characterise the historical economic growth.

Galor explains the mechanism of Malthusian stagnation but there was no stagnation. In its place there was a steadily-increasing and unconstrained hyperbolic growth. What needs to be explained is why the economic growth was hyperbolic but this feature is totally unrecognised in the Unified Growth Theory.

Galor explains the transition from stagnation to growth but this transition never happened because there was always a steadily-increasing and undisturbed economic growth. What needs to be explained is why at a certain stage, and relatively recently, the economic growth in various regions was diverted to slower trajectories (Nielsen, 2015c). This common feature is also unrecognised in the Unified Growth Theory.

Unified Growth Theory is repeatedly contradicted by Maddison's data (Maddison, 2001, 2010). Paradoxically, it is contradicted by the same data, which were used (but not analysed) during the formulation of this theory. A new theory would have to be in agreement with these data.

Unified Growth Theory is contradicted by the GDP data describing the world economic growth and the growth in Western Europe (Nielsen, 2014). It is contradicted by the GDP/cap data (Nielsen, 2015a). It is contradicted by the data for Africa (Nielsen, 2015b). It is contradicted by the economic growth in Asia (Nielsen, 2015d) and now it is contradicted by the data for the former USSR. However, implicitly, this theory is also contradicted by the extensive mathematical analysis of the economic growth in various regions and countries (Nielsen, 2015c) showing that the historical economic growth was hyperbolic.

While it is true that hyperbolic distributions can be hard to understand, it is also true that their analysis is so simple that it is trivial when using the reciprocal values of data (Nielsen, 2014). The GDP/cap distributions are even more puzzling and confusing but in principle their analysis is also simple (Nielsen, 2015a). The GDP or GDP/cap data have to be analysed with care. Mutilating them is not helpful (Ashraf, 2009; Galor, 2005a, 2005b, 2007, 2008a, 2008b, 2008c, 2010, 2011, 2012a, 2012b, 2012c; Galor and Moav, 2002; Snowdon & Galor, 2008). Such mutilations of data can be hardly expected to lead to reliable conclusions. On the contrary they lead readily to the incorrect concepts, interpretations and explanations. They lead to many "Mysteries of the growth process" (Galor, 2005a, p. 220), the mysteries, which have no connection to the real world.

Unified Growth Theory needs to be revised but it would have to be revised by carrying out mathematical analysis of Maddison's data (Maddison, 2001, 2010). Most likely, however, a new theory would have to be proposed, maybe even by Galor who devoted 20 years of his life (Baum, 2011) to develop his theory.

Here is a new and open field of research for economists, a field that will lead to a better interpretation of the mechanism of the economic growth. The new theory would have to abandon the current incorrect, misleading and potentially dangerous concept that after ages-long epoch of stagnation we have finally escaped the Malthusian trap and entered into a new epoch of prosperity supported by the sustained economic growth. The past economic growth was sustained but it was also slow and secure. The current economic growth is still sustained but it is fast and insecure (Nielsen, 2015e, 2015f).

We have not escaped the Malthusian trap because there was no trap, and we have not entered a new regime of sustained economic growth with the vision of the increasing prosperity. Our current and fast economic growth is not the reason for celebrations but for concern.